\documentclass[reprint,prl,showpacs]{revtex4-1}
\usepackage{amsmath,amssymb,bm}
\usepackage{graphicx}
\usepackage{hyperref}

\begin{document}
\title{Counting Pseudo Landau Levels in Spatially Modulated Dirac Systems}
\date{\today}
\author{Toshikaze Kariyado}
\affiliation{International Center for Materials Nanoarchitectonics (WPI-MANA),
National Institute for Materials Science, Tsukuba 305-0044, Japan}
\email{kariyado.toshikaze@nims.go.jp}

   \begin{abstract}
    In a system with Dirac cones, spatial
    modulation in material parameters induces a \textit{pseudo}
    magnetic field, which acts like an external magnetic field. Here,
    we derive a concise formula to count the \textit{pseudo} Landau
    levels in the simplest setup for having a pseudo magnetic field. The
    formula is
    so concise that it is helpful in seeing the essence of the
    phenomenon, and in considering the
    experimental design for the pseudo magnetic field. Furthermore, it
    is revealed that anisotropic Dirac cones are advantageous in pseudo
    Landau level formation in general. The proposed setup is relatively
    easy to be realized by spatial
    modulation in the chemical composition, and we perform an
    estimation of the pseudo
    magnetic field in an existing material (an antiperovskite material),
    by following the composition dependence with the help of
    the ab-initio method. 
   \end{abstract}

 \maketitle

 The external magnetic field is not the only source of the Landau
 levels. For instance, a certain strain on graphene leads to the
 so-called pseudo magnetic field and the resultant Landau
 level structures
 \cite{Guinea:2010aa,doi:10.1021/nl1018063,Levy544,YEH20111649}. This
 phenomenon is tied to the most intriguing
 property of graphene, i.e., the emergent relativistic electron, or the
 existence of Dirac cone in the band structure
 \cite{RevModPhys.81.109}. Having a Dirac cone,
 the key toward the finite pseudo magnetic field is the resemblance
 between the shift of the Dirac cone in the Brillouin zone and the
 minimal coupling in the U(1) gauge theory. Since the essence is simply
 the Dirac cone shift, the idea is not limited to graphene, but is
 applicable to any system with emergent linear dispersion, such as
 three-dimensional Dirac/Weyl semimetals \cite{PhysRevB.87.235306,PhysRevB.88.125105,PhysRevB.89.081407,PhysRevLett.115.177202,PhysRevX.6.041021,PhysRevX.6.041046,PhysRevB.94.241405,PhysRevB.95.041201,PhysRevB.95.115422,PhysRevB.95.125306}.

 The study of the pseudo magnetic field has several important
 aspects. Obviously, it is conceptually interesting to observe magnetic
 phenomena like chiral magnetic effect
 \cite{PhysRevB.94.241405} or quantum oscillations
 \cite{PhysRevB.95.041201} without actually applying magnetic
 field. Furthermore, the field strength can possibly exceed the maximum
 available strength for the real magnetic field. As an extreme case,
 even for a system inert to the real magnetic field (e.g. neutral
 particle systems, photonic or phononic crystals
 \cite{Brendel25042017,arXiv:1610.06406}), the pseudo magnetic field can
 be influential as far as there are Dirac
 cones. Typically, Dirac cones come with pairs,
 resulting in multiple Dirac
 nodes in the Brillouin zone, and the direction of the pseudo magnetic
 field depends on the nodes. (So, a pseudo magnetic field is regarded
 as an axial magnetic field.) Then, if the real and pseudo magnetic
 field coexist,
 they enhance or cancel with each other depending on the nodes, which
 induces the valley imbalance \cite{PhysRevB.82.205430,PhysRevB.87.121408,PhysRevB.95.041201,1705.09085}. In that sense, the study of the pseudo
 magnetic field also has potential importance in valleytronics as next
 functionalization of materials.

\begin{figure}[tbp]
 \begin{center}
  \includegraphics{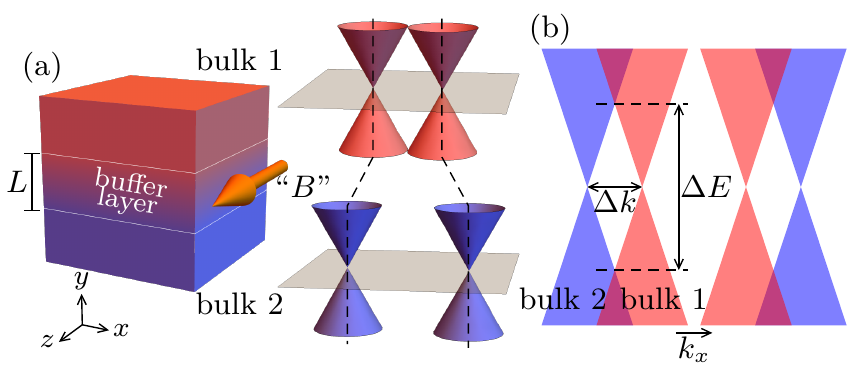}
  \caption{(a) (Color online) Setup under consideration. Bulk 1 and bulk
  2 are similar
  to each other, but with some relative shift of the Dirac cones in
  $k_x$ direction. ``$B$'' represents the direction of the
  pseudo magnetic field. (b) Bulk contribution to the band structure as a
  function of $k_x$.}\label{fig:setup}
 \end{center}
\end{figure}
As we have noted, the essence of the pseudo magnetic field is the
similarity between the Dirac/Weyl node shift and the minimal coupling. 
The shift of Dirac/Weyl nodes can be
induced in many ways \cite{Montambaux2009,PhysRevB.88.245126,Kariyado:2015aa}, and the
strain has been frequently used in the context of the pseudo magnetic
field in the literature. However, the strain is not the only
choice: the spatial dependence of the magnetic moment
\cite{PhysRevB.87.235306,PhysRevB.94.115312}, or the spatially
modulating chemical composition should be equally sufficient. In this
paper, we consider a
simple setup for the pseudo magnetic field generation, which is expected
to be relatively easy to realize with the spatial modulation of the
chemical composition.
We first give a notably concise formula [Eq.~\eqref{eq:master}] for the
number of observable pseudo Landau levels. The formula requires only two
dimensionless parameters $N$ and $R$, where $N$ characterizes the length
scale of the spatial modulation, while $R$ characterizes the size of the
Dirac/Weyl node shift. The conciseness of the formula makes the essence
of the pseudo magnetic field transparent, and helps to consider 
experimental designs. Furthermore, inspired by the simple formula, it is
pointed out that anisotropic Dirac cones are better than the isotropic
ones to appreciate the pseudo Landau level structure. 
In the latter half of this paper, we perform semi ab-initio estimation
of $R$ in a real material, antiperovskite A$_3$SnO (A=Ca,Sr)
\cite{JPSJ.80.083704,JPSJ.81.064701,klintenberg,PhysRevB.90.081112}, to
make a quantitative
argument on the pseudo magnetic field generation.

Let us start with the formulation. The setup in this paper
is illustrated in Fig.~\ref{fig:setup}(a). The considered system
consists of three regions, bulk 1, bulk 2, and the buffer region. For
having a transparent discussion, we assume that bulk 1 (bulk 2)
extends to $y=+\infty$ ($y=-\infty$), which excludes free surfaces from
our consideration. Bulk 1
and bulk 2 are similar to each other, having Dirac
cones in the band structure, but with slightly different node positions.
Namely, the effective model for each region is 
\begin{equation}
 H^{(\pm)}_{\bm{k}}=\hbar{v}(\bm{k}\pm\bm{k}_0)\cdot\bm{\sigma},\label{eq:H0}
\end{equation}
where $\bm{k}_0$, which denotes the center of Dirac cones, takes
different values for the two regions [see Fig.~\ref{fig:setup}(a)].
In the buffer region, we assume that the
Dirac cones are smoothly shifted from the position in bulk 1 to the one
in bulk 2. For simplicity, our focus is limited to the case that the
Dirac cones are shifted in $k_x$ direction. As we can see from
Fig.~\ref{fig:setup}(a), the system is periodic in $x$ and $z$
direction, and we have a band structure as a function of $k_x$ and
$k_z$. (If a two-dimensional Dirac/Weyl system is our target, we simply
neglect any structure along $z$-axis, and omit $k_z$.) Then, the bulk
contribution to the band structure typically looks like
Fig~\ref{fig:setup}(b) reflecting the shifted node positions. The
question is, how about the contribution from the state at the buffer
region.

Within the buffer region, the position of the Dirac cones depends on
$y$. Then, in a naive treatment, the effective model is assumed to be
\begin{equation}
 H^{(\pm)}=\hbar{v}(-i\bm{\nabla}\pm \bm{k}_0(y))\cdot\bm{\sigma},\label{eq:H1}
\end{equation}
which is obtained by replacing $\bm{k}$ by $-i\bm{\nabla}$ and taking
account of the $y$ dependence of $\bm{k}_0$. The latter arguments reveal
that this naive treatment works well. By comparing Eq.~\eqref{eq:H1}
with the standard minimal coupling $(-i\bm{\nabla}-e\bm{A})$, we have
\begin{equation}
 \bm{A}^{(\pm)}=\mp \frac{\hbar}{e}\bm{k}_0(y).
\end{equation}
Once $\bm{A}$ is given, the pseudo magnetic field is given by
$\bm{B}=\bm{\nabla}\times\bm{A}$. In our setup, the node shift is only
in $k_x$ direction, and depends only on $y$, which means that 
$\frac{\partial A_x}{\partial y}$ is the only relevant
component. Assuming that $\bm{k}_0$ linearly interpolates bulk 1 and bulk
2, the pseudo magnetic field strength $B=|\bm{B}^{(\pm)}|$ is estimated as
\begin{equation}
 B\sim \frac{\hbar}{e}\frac{\Delta k}{L},\label{eq:B0}
\end{equation}
where $L$ is the thickness of the buffer region, and $\Delta k$ is the
size of the node shift [See Fig.~\ref{fig:setup}(b)]. 

To make the formulation concise, we introduce two dimensionless
parameters $N$ and $R$ respectively as $L=Na$ and $\Delta k=2\pi R/a$,
where $a$ is the lattice constant. (It is implicitly assumed that the
lattice constants in $x$ and $y$ are the same, but the extension to 
anisotropic cases is trivial.) $N$ represents the length scale of the
spatial modulation, while $R$ measures the size of the node shift in the
unit of the Brillouin zone size. As a rough estimation, using a typical
atomic scale $a\sim 5$ \AA, Eq.~\eqref{eq:B0} leads to
\begin{equation}
 B \sim 1.6\times 10^4\times \frac{R}{N}\text{ [T]}.\label{eq:B1}
\end{equation}
That is, we potentially have 16 thousand Tesla, and the available
strength is reduced by a factor of $R/N$.

\begin{figure}[tbp]
 \begin{center}
  \includegraphics{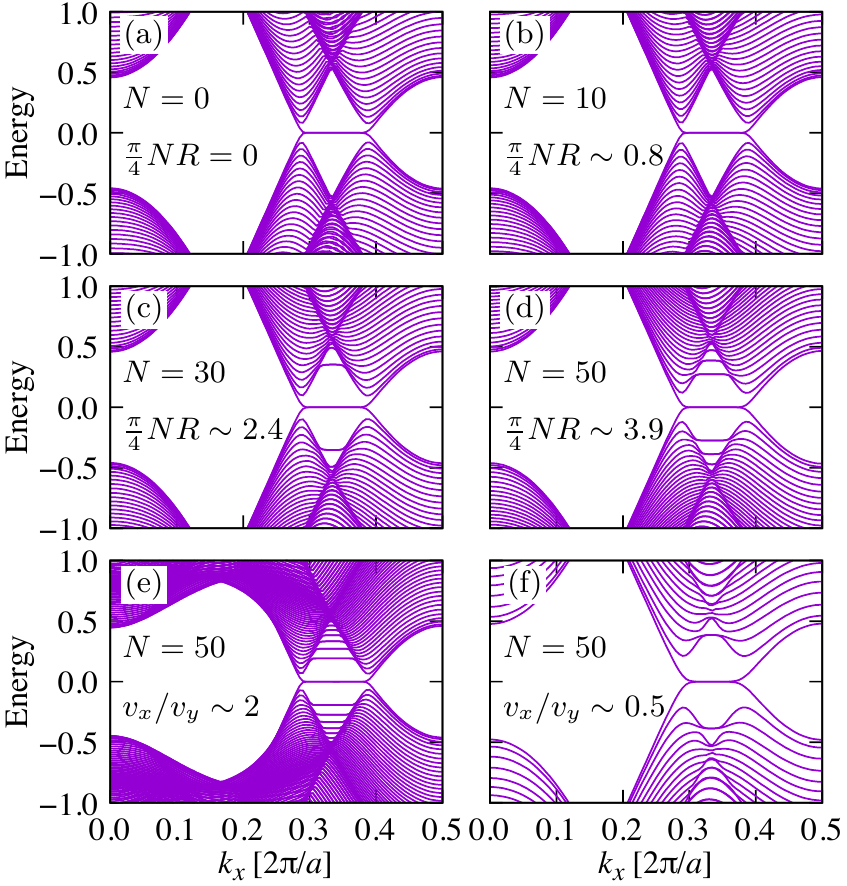}
  \caption{(Color online) The band structure with the pseudo magnetic field for several values of $N$. For (e) and (f), the anisotropy of the Dirac cone is
  introduced.}\label{fig:toy}  
 \end{center}
\end{figure}

The obtained pseudo magnetic field induces the Landau levels. Plugging
Eq.~\eqref{eq:B0} into a textbook formula, the energy of $n$th Landau level
is 
\begin{equation}
 E_n=\pm\sqrt{\frac{4\pi v^2\hbar^2R|n|}{Na^2}}. \label{eq:En}
\end{equation}
[For three-dimensional systems, Eq.~\eqref{eq:En} corresponds to the
energy at $k_z=0$.]
However, we should note that not the all Landau levels are observable in
the energy spectrum. That is, since we have the bulk regions as well as
the buffer region, the Landau levels in the buffer region can be masked
by the bulk contribution in the energy spectrum. It turns out that the
diamond region with height $\Delta E$ and width $\Delta k$ in
Fig.~\ref{fig:setup}(b) is available for the Landau levels (see also the
latter arguments on the numerical results in Fig.~\ref{fig:toy}).
Since $\Delta E$ is estimated as $\Delta E=\hbar v\Delta k$, the condition
that the $n$th Landau level falls into this diamond region becomes
\begin{equation}
 \sqrt{\frac{4\pi v^2\hbar^2R|n|}{Na^2}}<\frac{\hbar v\Delta k}{2},\label{eq:compare}
\end{equation}
which leads to a concise expression
\begin{equation}
 |n|<\frac{\pi}{4}NR.\label{eq:master}
\end{equation}
Here we make a short summary: (i) to make the pseudo magnetic field
strong, $N$ should be small [Eq.~\eqref{eq:B1}], (ii) to observe the large
number of Landau levels, $N$ should be large [Eq.~\eqref{eq:master}],
and (iii) large $R$ is always beneficial.

Let us move on to the numerical validation of the derived formula. For
this purpose, we introduce a two-dimensional square lattice
tight-binding model with mobile Dirac nodes. Specifically, the
Hamiltonian is
\begin{equation}
 H_{\bm{k}}=[1+\delta+2(\cos k_x+\cos k_y)]\sigma_z
  + 2\alpha \sin k_y \sigma_y
\end{equation}
where the lattice constant $a$ is set to $1$. 
By expanding the Hamiltonian with respect to
$\tilde k_x\equiv k_x-2\pi/3$ and $k_y$ up to the first order in each of
the parameters, we end up with
\begin{equation}
 H_{\bm{k}}\sim -\sqrt{3}
  \bigl[
  (\tilde{k}_x-\tilde{\delta})\sigma_z+\tilde{\alpha} k_y\sigma_y
\bigr],
\end{equation}
where $\tilde{\delta}=\delta/\sqrt{3}$ and
$\tilde{\alpha}=2\alpha/\sqrt{3}$. That is, $\tilde{\delta}$ behaves as
$A_x$, and $\tilde{\alpha}$ is essentially anisotropy of the Fermi
velocity, $v_y/v_x$. Therefore, if we
assign $\tilde{\delta}/2\pi=0.05$ for bulk 1,
$\tilde{\delta}/2\pi=-0.05$ for bulk 2, and the linearly
interpolated value for the buffer region, $R=0.1$ is achieved. 
In the actual calculation, we make the system periodic also in $y$
direction not to have free edges. Namely, the system consists of the
repetition of the chunk of bulk 1 -- buffer -- bulk 2 -- buffer. 

Figures \ref{fig:toy}(a)--\ref{fig:toy}(d) summarizes the results for
$\tilde{\alpha}=1$. If there appears some flat sections in the band
structure, they can be regarded as the Landau levels. (A flat
section gives a peak in the density of state.)
For $N=0$
[Fig.~\ref{fig:toy}(a)], i.e., if the
change between bulk 1 and bulk 2 is sharp, only the zeroth Landau level
is identified in the band structure. For $N=1$ [Fig.~\ref{fig:toy}(b)],
which results in $\frac{\pi}{4}NR\sim 0.8$, we still only see the zeroth
Landau level. If $N$ is further increased to $\frac{\pi}{4}NR\sim 2.4$
[Fig.~\ref{fig:toy}(c)], the $n=1$ Landau level becomes clearly visible,
and we see a small signature of the $n=2$ Landau level as well. For
$N=50$ leading to $\frac{\pi}{4}NR=3.9$ [Fig.~\ref{fig:toy}(d)], the
clear identification of the Landau levels up to $n=3$ is possible. All
of these observations confirm the formula Eq.~\eqref{eq:master}.

It is worth noting that the expected peak structure in the (local)
density of state should be a key to experimental detection of the pseudo
Landau levels. Any measurements capable of detecting the density of
state, such as STM/STS as a direct measurement or optical conductivity,
might be useful. 

So far, we have been treating the isotropic Dirac cone. Actually, the
anisotropy of the Dirac cone gives significant influence on the
observable Landau levels. If the Dirac cone becomes anisotropic, $v^2$
in the left hand side of Eq.~\eqref{eq:compare} is replaced by $v_xv_y$,
and $v$ in the right hand side by $v_x$. Consequently, the formula
Eq.~\eqref{eq:master} is rewritten as 
\begin{equation}
 |n|<\frac{\pi}{4}\frac{v_x}{v_y}NR.\label{eq:aniso}
\end{equation}
This implies that if we have $v_x>v_y$, the number of observable Landau
levels increases compared with the isotropic case with the same $NR$. 
Physically, this is because larger $v_x$ means
larger $\Delta E$, and smaller $v_y$ means larger density of
states, both of which is advantageous to observe more Landau levels. 
The formula Eq.~\eqref{eq:aniso} is again confirmed using the toy model
by modifying $\tilde{\alpha}$. Figures~\ref{fig:toy}(e) and
\ref{fig:toy}(f) shows that for $v_x/v_y\sim 2$ ($\tilde{\alpha}=2$),
the number of the Landau levels is doubled comparing with the isotropic
case, while for $v_x/v_y\sim 0.5$ ($\tilde{\alpha}=0.5$), only the $n=0$ and
$n=1$ Landau levels are clearly seen. Therefore, if one attempts
to observe large number of pseudo Landau levels, it is better to focus
on the system with anisotropic Dirac cones. 

Hereafter, we work on the quantitative estimation of the pseudo magnetic
field in existing materials. Having formulae Eqs.~\eqref{eq:master} and
\eqref{eq:aniso}, the estimation of $R$ is essential, and we derive $R$ in
semi ab-initio
way. Here, semi ab-initio means that we apply the first-principles
density functional theory \cite{QE-2009} with a small assumption in the
crystal structure to calculate the electronic band structure. 
Since the two-dimensional cases have been studied in graphene
extensively, our focus is on the three-dimensional cases -- we take the
cubic antiperovskite family A$_3$SnO (A=Ca,Sr), where the
three-dimensional Dirac cones are found on $k_x$, $k_y$, and $k_z$ axes
in the first-principles calculation. (Note that the experimental studies
on this materials are now in progress
\cite{doi:10.1063/1.4820770,Nuss:dk5032,doi:10.1063/1.4955213,doi:10.1063/1.4952393,Oudah:2016aa}.)
The great advantage of the
antiperovskite family is that we already know the way to tune the
electronic structure
near the Fermi energy at the qualitative level \cite{twin}. Namely, it
is natural
to expect that the Dirac cone shift is realized by preparing
Ca$_{3(1-x)}$Sr$_{3x}$SnO and adjusting $x$ \cite{twin}. Before we proceed, we
would like
to point out a minor disadvantage of the antiperovskite family. Strictly
speaking, there is a tiny mass gap at the Fermi energy, and therefore, we
have to achieve $R$ such that $\Delta E$ is sufficiently larger than the
mass gap. 

\begin{figure}[tbp]
 \begin{center}
  \includegraphics{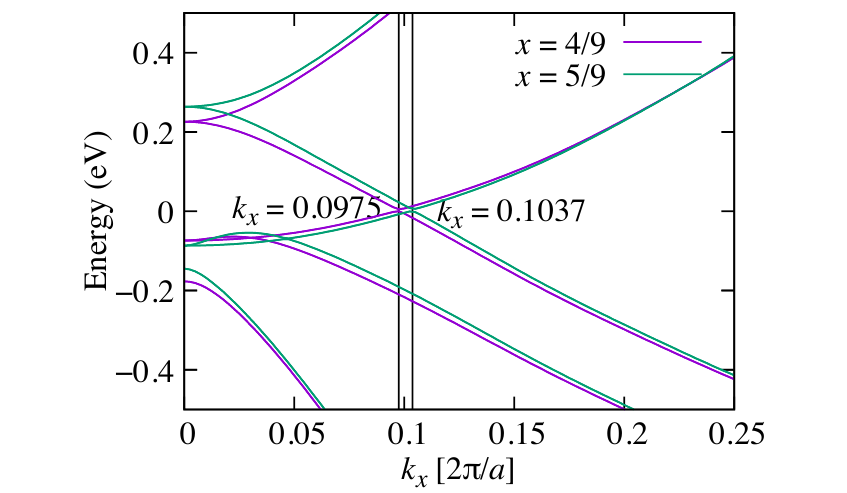}
  \caption{(Color online) The band structure of
  Ca$_{3(1-x)}$Sr$_{3x}$SnO obtained with the
  interpolation using the MLWF method.}\label{fig:wannier}
 \end{center}
\end{figure}
Here, the electronic structure for $0<x<1$ is obtained in the following
way. Firstly, we apply the first-principles calculation for $x=0$ and
$x=1$. (The computational details are in parallel with Ref.~\onlinecite{twin}.)
Using those results, we construct effective
models for $x=0$ and $x=1$, and then, the model for arbitrary $x$ is
obtained by interpolating the parameters in the effective model. To be
quantitative, the construction of the effective model is conducted using
the maximally localized Wannier function method implemented in Wannier90
package \cite{MOSTOFI2008685}. In practice, we construct a 12 orbital
model where \textit{12} comes from (\textit{2} spins) $\times$
(\textit{3} $p$-orbitals on Sn atom + $d_{x^2-y^2}$/$d_{y^2-z^2}$/$d_{z^2-x^2}$ on \textit{3} crystallographically
equivalent Ca/Sr atoms) $=$\textit{2}$\times$(\textit{3}+\textit{3})
\cite{JPSJ.81.064701}. Since we intend to focus on the
small variation around $x=0.5$, we
fix the lattice constant as the average of the lattice constants for
$x=0$ and $x=1$, which are experimentally known, throughout the
calculation. The band structures along $k_x$ axis for $x=4/9$ and
$x=5/9$ obtained by the
interpolation are shown in Fig.~\ref{fig:wannier}. The crossing points
at the Fermi energy are the Dirac cones in this system. The inspection
of the band structure reveals that the Dirac cone locates at $k_x\sim 0.0975$
for $x=4/9$, while $k_x\sim 0.1037$ for $x=5/9$, where the momentum is
measured in the unit of $2\pi/a$. At the end, $R$ is estimated to be
approximately 0.006.

\begin{figure}[tbp]
 \begin{center}
  \includegraphics{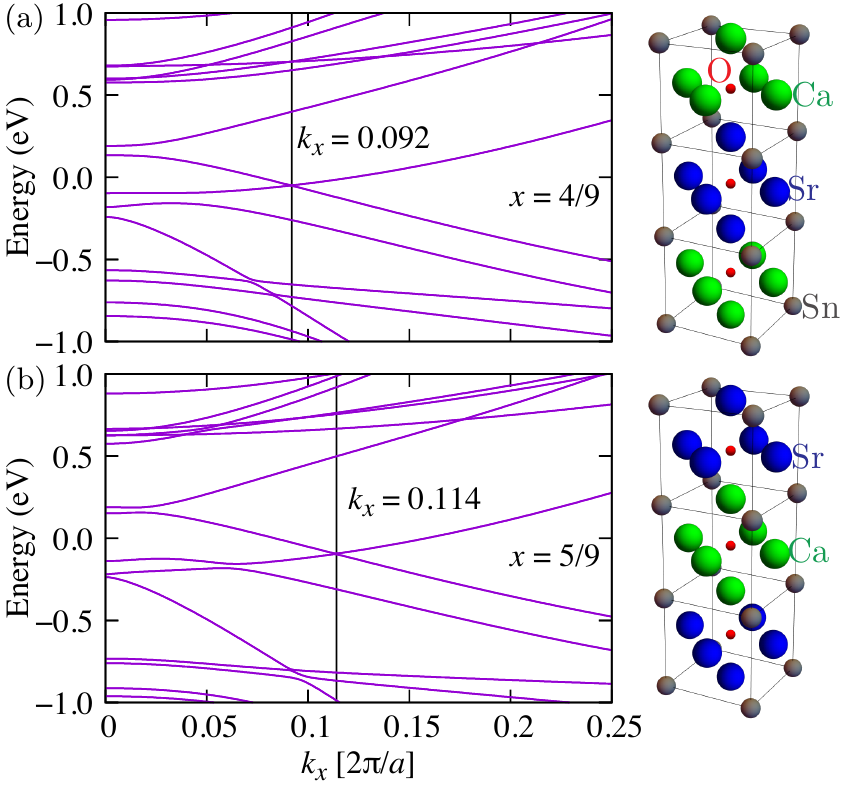}
  \caption{(Color online) The band structure calculated using the
  enlarged unit cell
  with the composition Ca$_{3(1-x)}$Sr$_{3x}$SnO. The used crystal structures
  are shown in the right panels.}\label{fig:hetero}
 \end{center}
\end{figure}

As a study complementary to the Wannier orbital based interpolation, we
perform the
calculation with a superstructure, containing Ca and Sr in a ratio
$x=4/9$ and $x=5/9$. In specific, we consider the superstructure in $z$
direction as shown in the right panels of
Fig.~\ref{fig:hetero}. Since the unit cell is not enlarged in $x$-$y$
plane, the node shift in $k_x$ direction can be discussed in exactly the
same footing as the previous paragraph (no Brillouin zone folding in
$k_x$ and $k_y$ direction). Again, assuming that $x=4/9$ and $x=5/9$ are
sufficiently close to $x=0.5$, we use the averaged lattice constant. 
The band structures for $x=4/9$ and $x=5/9$ on the $k_x$ axis are shown
in Fig.~\ref{fig:hetero}. From this result, we extract the node position for
$x=4/9$ as $k_x=0.092$, while the position for $x=5/9$ as $k_x=0.114$,
which leads to $R=0.022$.

To summarize, we derive a compact formula to count the number of the
observable Landau levels induced by the pseudo magnetic field. 
The formula is so simple that the essence of the pseudo magnetic field
becomes evident. Having a concise formula is also beneficial in
designing experiments on the pseudo Landau levels in real materials. In
fact, we show an explicit estimation of the pseudo magnetic field in an
antiperovskite Dirac material in an ab-initio manner. For future
developments, it is pointed out that anisotropic Dirac cones are more
advantageous than isotropic ones in appreciating the pseudo Landau level
structure. 

\begin{acknowledgments}
 The author thanks M.~Ogata, H.~Takagi and A.~Vishwanath for stimulating
 discussions. The computation in this work has been done using the
 facilities of the Supercomputer Center, the Institute for Solid State
 Physics, the University of Tokyo. This work was supported by a
 Grant-in-Aid for Scientific Research No.~17K14358.
\end{acknowledgments}

\end{document}